\documentclass[3p,times]{elsarticle}

\usepackage{ecrc}

\usepackage{epstopdf}

\volume{00}

\firstpage{1}

\journalname{Nuclear Physics A}

\runauth{Romuald A. Janik}

\jid{nupha}

\jnltitlelogo{Nuclear Physics A}

\usepackage{graphicx}
\usepackage{amsmath,amssymb}

\newcommand{\red}[1]{{#1}}
\newcommand{\green}[1]{{#1}}

\newcommand{\black}[1]{{#1}}
\newcommand{\blue}[1]{{#1}}
\newcommand{\violet}[1]{{#1}}

\usepackage[utf8]{inputenc}

\newcommand{\eq}{\begin{equation}}
\newcommand{\eqx}{\end{equation}}
\newcommand{\eqn}{\begin{eqnarray*}}
\newcommand{\eqnx}{\end{eqnarray*}}

\newcommand{\f}[2]{\frac{#1}{#2}}

\newcommand{\eps}{\varepsilon}
\newcommand{\al}{\alpha}
\newcommand{\bt}{\beta}

\newcommand{\dl}{\delta}
\newcommand{\Dl}{\Delta}
\newcommand{\sg}{\sigma}
\newcommand{\om}{\omega}

\newcommand{\teff}{T_{eff}}

\newcommand{\nn}{{\cal N}}

\newcommand{\oo}[1]{{\cal O}\left(#1\right)}

\newcommand{\qqqq}{\quad\quad\quad\quad}
\newcommand{\qq}{\quad\quad}

\begin{document}

\begin{frontmatter}



\title{AdS/CFT for the early stages of heavy ion collisions}

\author{Romuald A. Janik}
\address{Institute of Physics\\
Jagiellonian University\\
ul. Reymonta 4, 30-059 Kraków, Poland}




\begin{abstract}
I give a brief introduction to the AdS/CFT correspondence targeted at heavy-ion physicists. I also review some insights to our
understanding of the early stages of heavy-ion collisions coming from selected studies made using methods of the AdS/CFT
correspondence.
\end{abstract}

\begin{keyword}
AdS/CFT correspondence \sep quark-gluon plasma \sep thermalization

\end{keyword}

\end{frontmatter}



\section{Introduction}

In this talk I will describe recent understanding of the physics of early stages of heavy-ion collisions coming from
the AdS/CFT correspondence \cite{ref1}. Since the AdS/CFT correspondence may be still rather exotic for the heavy-ion
community, I will spend the first part of this talk describing the physical content of the correspondence and what kind
of physics of strongly coupled plasma is naturally contained in the dual AdS/CFT description. Then I will move on
to describe some selected results obtained using the methods of the AdS/CFT correspondence in the course of the last
two years since the previous Quark Matter conference. I will naturally not give any technical details but concentrate
on the physical insights obtained.

\section{The AdS/CFT correspondence}

The AdS/CFT correspondence states the equivalence of two apparently completely unrelated theories: a four
dimensional $\nn = 4$ supersymmetric Yang-Mills theory and superstring theory on a specific, curved ten dimensional
background: $AdS_5\times  S^5$. The gauge theory contains gluons and, in addition, 6 scalar and 4 fermion fields in the
adjoint representation. This matter content makes the theory completely different from QCD at zero temperature,
however when one considers both theories at sufficiently high temperature (on the QCD side above the deconfinement
temperature with some margin), the physics of both theories becomes qualitatively quite similar. In fact, a recent study
of various perturbative transport and collisional properties etc. indicates that the $\nn = 4$ Super Yang-Mills plasma is
very similar to QCD plasma with the main differences reflecting the different number of degrees of freedom of the
two theories \cite{ref2}.

On the nonperturbative level, there are both similarities and differences. Both plasma systems are of course
deconfined and strongly coupled. Moreover both of them have no supersymmetry. This point is worth emphasizing,
as $\nn = 4$ SYM is in fact supersymmetric. However, once we consider the theory at some nonzero temperature, or
more generally in a state with some nonzero energy density, supersymmetry is broken and the various miraculous
cancellations which caused the $\nn = 4$ SYM to be so drastically different from QCD in the vacuum disappear. Having
said that, one has to nevertheless keep in mind some differences, which may be relevant or not depending on the
physics being addressed. Firstly, in $\nn = 4$ SYM there is no running coupling, so even at very high energy densities
the coupling remains strong. This may not be a big problem for bulk properties but may be significant for observables
which involve a mixture of perturbative and nonperturbative physics. Secondly, there is an exactly conformal equation
of state (i.e. $E = 3p$ or more generally $T_{\mu}^\mu = 0$). In particular the bulk viscosity vanishes. This is certainly not a
good approximation close to $T_c$ . However for higher temperatures this may be reasonable depending on the desired
precision. Finally, in the $\nn = 4$ theory, there is no confinement/deconfinement transition, so the plasma expands and
cools indefinitely. This is the biggest difference to keep in mind.

There are more refined (and more complicated) versions of the AdS/CFT correspondence which allow us to lift
some of these differences, however, the investigations related to the early stage of heavy-ion collisions involved the
AdS/CFT correspondence in its simplest setting dealing with $\nn = 4$ SYM.

\section{The physical meaning of the AdS/CFT correspondence}

The AdS/CFT correspondence postulates the equivalence of the 4-dimensional gauge theory with superstring
theory in $AdS_5 \times S^5$. Thus the string theory here is not some exotic physics but rather an extremely non-obvious
repackaging of gauge theory degrees of freedom. This is especially interesting from the perspective of heavy-ion
physics, as some of these degrees of freedom behave (in some cases) like collective hydrodynamic excitations of
the plasma. Moreover the question whether these particular degrees of freedom are dominant or not is a genuine
dynamical question which can be answered by concrete computations.

More generally, the dual string degrees of freedom contain both gravity modes and massive string modes. The
utility of the AdS/CFT correspondence stems from the fact that at strong coupling in gauge theory, 
there appears a huge mass gap
between the gravity and massive string modes, so the latter ones decouple and the dynamics can be reduced to gravity
described by Einstein’s equations. At weak coupling in gauge theory, all those modes are intermingled and we do not have a workable
AdS/CFT description for plasma physics.

\subsection*{What physics is captured by the AdS/CFT description?}

To get some intuition about the physics captured by the AdS/CFT description it is interesting to consider the
excitations of an infinite uniform static plasma system. The uniform plasma is described on the dual gravity side by a
planar black hole. Small disturbances of the plasma are described by small perturbations of the 5-dimensional black hole metric
\eq
g^{5D}_{\al\bt}=g^{5D,black\ hole}_{\al\bt}+\dl g^{5D}_{\al\bt}(z) \,
e^{-i\red{\om} t+i \red{k} x} 
\eqx
where $z$ is the $5^{th}$ AdS coordinate. On the gauge theory side this gets directly translated to plane wave perturbations
of the energy-momentum tensor
\eq
T_{\mu\nu}=T_{\mu\nu}^{static}+ t_{\mu\nu}\, e^{-i\red{\om} t+i \red{k} x}
\eqx
Imposing ingoing boundary conditions at the black hole horizon for given $k$ fixes a discrete set of frequencies $\om$ (these
are the so-called quasinormal modes (QNM) – see \cite{ref4} for more details). Changing $k$ provides a dispersion relation for
these modes. It turns out that the lowest modes exactly coincide with hydrodynamic shear and sound modes:
\eq
\!\!\!\!\!\red{\om_{shear}}=-i \f{\eta}{E+p} \red{k^2} + \oo{\red{k^3}} \qqqq
\red{\om_{sound}}=\f{1}{\sqrt{3}} \red{k} -i \f{2}{3} \f{\eta}{E+p} \red{k^2} + \oo{\red{k^3}}
\eqx
In addition, there is an infinite set of genuine nonhydrodynamical modes
\eq
\red{\om_{non-hydro}^{(n)}}= - i \Gamma_n \pm \Omega_n + \oo{\red{k^\#}}
\eqx
Each of these modes represents a different collective excitation of the strongly coupled gauge theory plasma. Close to
equilibrium they are damped and decay exponentially, but as we will see later, in the early stages of plasma evolution
their dynamics dominates over the hydrodynamic modes. At first it might seem strange that there are so many more
degrees of freedom than hydrodynamics, however in fact it is very natural. Indeed hydrodynamics describes essentially
just the local energy density. There is a mutlitude of ways to distribute this energy among various modes of the plasma.
Once we are far from equilibrium we have to take into account the redistribution of energy between the modes and
the nonhydrodynamic degrees of freedom can be understood as being responsible for that at strong coupling.

\subsection*{Why use AdS/CFT?}

The main motivation for using the AdS/CFT correspondence in the context of heavy-ion physics is that within the
same framework we may compute the evolution of a strongly coupled plasma system from some nonequilibrium state
to a state described by hydrodynamics. The results apply directly for the N = 4 SYM theory, but we may expect some
qualitative similarity with real evolution of QCD quark-gluon plasma. At the very least, these results provide for us
examples of what behaviour may happen so that we may test our theoretical prejudices. Also if some universal feature
arises we should consider it seriously for realistic plasma evolution. In some cases the AdS/CFT correspondence may
provide for us some ball-park figures, the prime example being the AdS/CFT value of $\eta/s$ \cite{ref3}.

\section{Recent developments}

After this brief introduction to the AdS/CFT correspondence, I will now describe some selected recent develop-
ments. I will start with another look on boost-invariant equilibration, where we may observe some features touched
upon previously, in particular the role of the nonhydrodynamic degrees of freedom and their incorporation into a
purely 4-dimensional extension of hydrodynamics. Then after mentioning developments in high order hydrodynam-
ics and thermalization in 2+1 dimensions, I will describe the recent lessons from the study of shock wave collisions in
AdS. Finally I will describe a hybrid AdS/realistic QGP study, where the AdS simulations are used to provide initial
conditions for subsequent realistic hydrodynamic/hadronization codes.

\subsection{Another look on boost-invariant equilibration}

In \cite{ref5,ref6}, the evolution of a strongly coupled boost-invariant plasma system was studied starting from diverse
choices of initial conditions. This was done under the additional assumption of homogenity in the transverse plane.
This system is particularly interesting as it exhibits rich far from equilibrium physics at the same time being easy to
describe and parametrize.

The symmetry assumptions determine all components of the energy-momentum tensor in terms of a single function
— the energy density at mid-rapidity $\eps(\tau) $. It is convenient, however, to factor out the dependence on the number of
colours and the specific number of degrees of freedom in $\nn=4$ SYM by using instead an effective temperature
defined by
\eq
\eps(\tau) = \f{3}{8} N_c^2 \pi^2 \violet{\teff^4(\tau)}
\eqx
which is the temperature of a thermal system with the same energy density as the instantaneous energy density  $\eps(\tau) $ at
proper time $\tau$. It is important to emphasize that this is just a book-keeping device and does not assume in any way
thermal equilibrium.

The papers \cite{ref5,ref6} normalized the solutions to have the same energy density at $\tau = 0$. Since this quantity is not ex-
perimentally (and even theoretically) accessible, it is useful to normalize the solutions to have the same hydrodynamic
late time behaviour
\eq
\pi \teff(\tau) \sim \f{1}{\tau^{\f{1}{3}}} \qq \text{\black{in the $\blue{\tau\to \infty}$ limit}}
\eqx
This was possible as for all initial conditions studied in \cite{ref5,ref6} a clear transition to a hydrodynamic description was
observed. The ‘1’ in the numerator fixes units.

The evolution of the effective temperature for the various initial conditions is shown in figure 1. We observe a clear
transition to hydrodynamics, and before that a predominance of nonhydrodynamic degrees of freedom (if dissipative
hydrodynamics, even of an arbitrarily high order would be the correct description, we would observe instead just a
single curve).

Secondly, at the transition to hydrodynamics the anisotropy of $T_{\mu\nu}$ is sizeable:
\eq
\Dl p_L \equiv 1-\f{p_L}{\eps/3} \sim 0.7
\eqx
This means that the transition to hydrodynamics (also dubbed as hydrodynamization) is distinct from true thermaliza-
tion. The subsequent isotropization occurs just through dissipative viscous effects and not through some exotic fast
nonequillibrium physics. The key insight from AdS/CFT is that early thermalization, understood in particular as the
isotropy of the diagonal components of $T_{\mu\nu}$ in the local rest frame, is not neccessary for the applicability of viscous
hydrodynamics.

\begin{figure}[t]
\centerline{
\includegraphics[height=6cm]{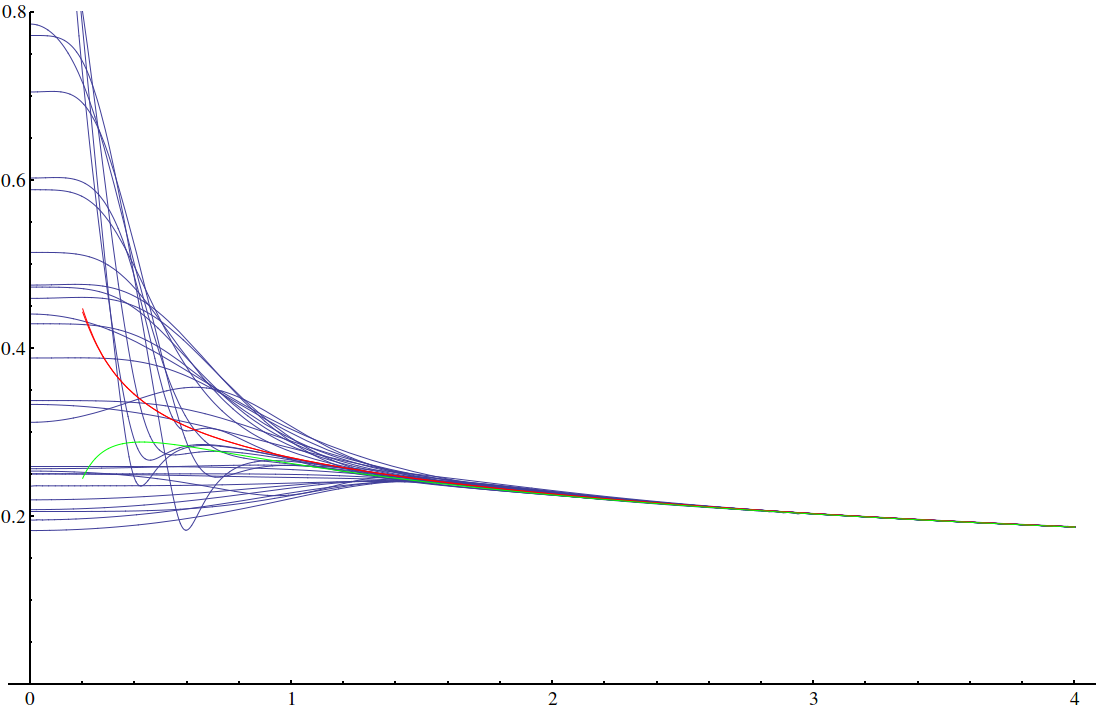}
}
\caption{
The evolution of the effective temperature $\teff$ as a function of proper time $\tau$ (in units defined by the asymptotics (6)) for various initial
conditions, obtained by solving numerically Einstein’s equations in the dual gravitational AdS/CFT description. The green and red curves (color online) represent, respectively, $3^{rd}$ order viscous hydrodynamics and Borel resummed dissipative hydrodynamics.}
\label{fig1}
\end{figure}

The final result of \cite{ref5,ref6} is that for the dimensionless product
\eq
w \equiv \teff \cdot \tau>0.6-0.7
\eqx
viscous hydrodynamics is applicable to a very high accuracy. For comparison, sample early initial conditions for
hydrodynamics at RHIC from \cite{ref7}, $T = 500\, MeV$ and $\tau_0 = 0.25\, fm$, give $w = 0.63$, so one might conclude that the
applicability of hydrodynamics is very natural at that point. It is reassuring that $\nn=4$ SYM provides an estimate
which is reasonable from the point of view of QCD phenomenology.

\subsection{Nonhydrodynamic degrees of freedom from a 4D perspective}

Given the importance of the nonhydrodynamic degrees of freedom before the onset of hydrodynamics and the
complexity of a full higher dimensional numerical relativity simulation, it is interesting to ask whether one can incor-
porate the lowest nonhydrodynamic degrees of freedom into a purely 4-dimensional effective description. Some ideas
about this have already been present in the literature \cite{ref8,ref8bis}, and such an extension of hydrodynamics is also in line with
other anisotropic versions of hydrodynamics \cite{ref9} which included additional degrees of freedom.
In \cite{ref10}, a proposal has been made on how to incorporate the dynamics of the lowest quasinormal mode in a way
which agrees with what we know from AdS/CFT.

The basic idea is to add new 4D quantities to the expression of $T_{\mu\nu}$ in terms of $T$ and $u^\mu$ and write equations of
motion for these additional degrees of freedom (these equations of course also depend on $u^\mu$ and $T$ ). 
Then the conservation law $\partial_\mu T^{\mu\nu}=0$
will couple these degrees of freedom back to the hydrodynamic ones. In the end we get equations which depend
on two additional parameters: the real and imaginary part of the quasinormal frequency $\om_R$ and $\om_I$ of the lowest
non-hydrodynamic mode.

\subsection{Some other developments}

Before discussing the most relevant case of shock wave collisions, let me mention two other recent developments.
We now have a very detailed understanding of high order dissipative hydrodynamics for strongly coupled $\nn=4$ SYM
in the boost-invariant context \cite{ref11}. Basically we have at our diposal about 240 terms in the late time expansion of the
effective temperature
\eq
\teff(\tau) = \f{1}{\tau^{\f{4}{3}}} \left[ 1+\sum_{\red{n}=1}^{\red{240}} \red{a_n} \f{1}{\tau^{\f{2\red{n}}{3}}} \right]
\eqx
with $a_n$ known numerically to a high accuracy. Apart from the study of some purely theoretical issues like (lack of)
convergence and Borel summability, such an expression may come to be useful when comparing various phenomeno-
logical higher order formulations of dissipative hydrodynamics.

Another interesting work was \cite{ref12}, where hydrodynamization of (inhomogeonous) UV sourced configurations in a
2+1 dimensional theory was investigated. The conclusion was that the evolution was well described initially by free
streaming which then went over to $2^{nd}$ order hydrodynamics. However, one has to keep in mind that the UV sourced
nature of the initial configuration was important here.

\begin{figure}[t]
\centerline{
\includegraphics[height=6cm]{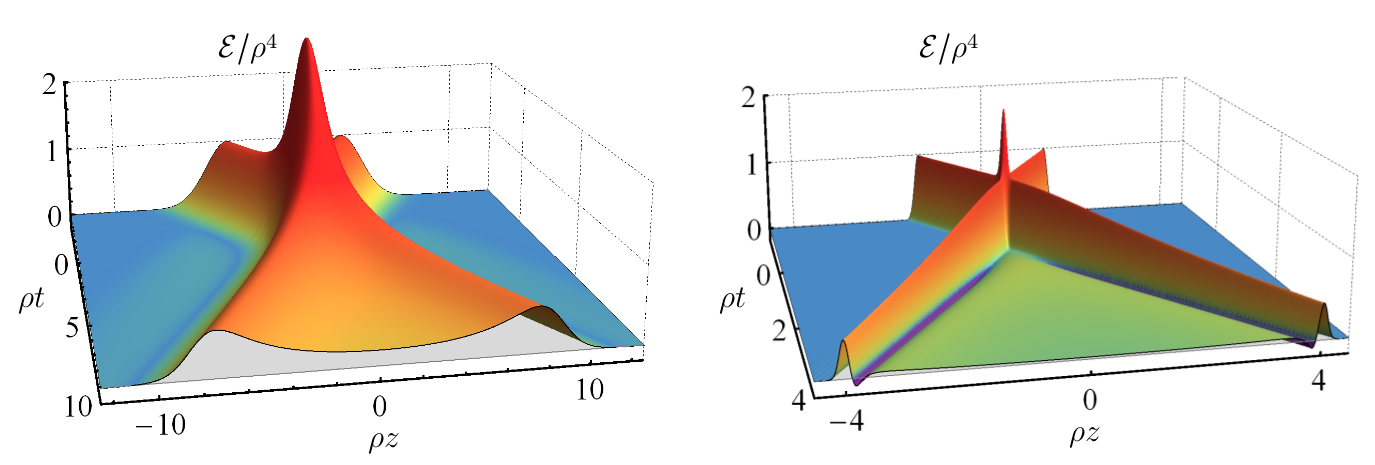}
}
\caption{
Energy profiles of thick and thin shock wave collisions (figure from \cite{ref17}).}
\label{fig2}
\end{figure}

\subsection{Shock wave collisions}

One problem that one has to face when attempting to use the AdS/CFT correspondence to model heavy-ion
collisions is what to choose as the projectile.

N = 4 SYM is different from QCD and we do not have a direct counterpart of the heavy ions which are being
collided. In \cite{ref13} it was suggested to use a shockwave which corresponds to a gauge theory state with the only
nonvanishing component of the energy-momentum tensor being
\eq
T_{--}=\mu\, \dl(x^-) \quad \left(\;\text{\black{or}} \quad T_{--}= f(x^-)\;\right)
\eqx
where $x^-=t-x$ is a light-cone coordinate.
The lack of dependence on transverse coordinates is again taken just for simplicity. These configurations have an
explicit and simple gravity dual and their collisions have been subject to a number of analytical explorations in various
approximations \cite{ref14,ref15}.

A pioneering numerical study was undertaken in \cite{ref16} with relatively thick shock waves (where the thickness is
measured by the extent in $x$ - relative to the energy density per transverse area). The main observation was that the
plasma system essentially stopped with the remnants moving away with velocity $v < c$.

In \cite{ref17,ref18,ref19} the problem of shock wave collisions was revisited uncovering quite complex behaviour. It turned
out that the character of the collision significantly depends on the thickness of the shock-wave (see figure 2). For
thick shock waves, full stopping and the behaviour observed in \cite{ref16} was recovered. In addition, it turns out that
hydrodynamics sets in very early, already practically at the peak of the energy density (see figure 2 on the left).

On the other hand, for thin shock waves, one observes transparency, the fragments move away with the speed
of light with regions of trailing negative energy density similar to the analytical picture of \cite{ref14}. At the transition to
hydrodynamics, \cite{ref17} observed a region of approximately constant (in the longitudinal coordinate $z$) 
energy density (in the center of mass frame). This translates to a Gaussian rapidity profile at mid-rapidity 
(more precisely $\sim (\cosh y)^{-\f{4}{3}}$).

In a subsequent paper \cite{ref18}, the authors observed a very interesting phenomenon of longitudinal coherence. They
studied superpositions of gaussian shock-waves with nontrivial longitudinal structure
\eq
T_{++}(x^+) = \f{N_c^2}{2\pi^2} \cdot \f{\violet{\mu^3}}{8\pi \green{\sg}}  \left\{
\exp\left[ -\f{(x^+ - \red{L/2})^2}{2\green{\sg^2}} \right] +
\exp\left[ -\f{(x^+ + \red{L/2})^2}{2\green{\sg^2}} \right]
\right\}
\eqx

\begin{figure}[t]
\centerline{
\includegraphics[height=5cm]{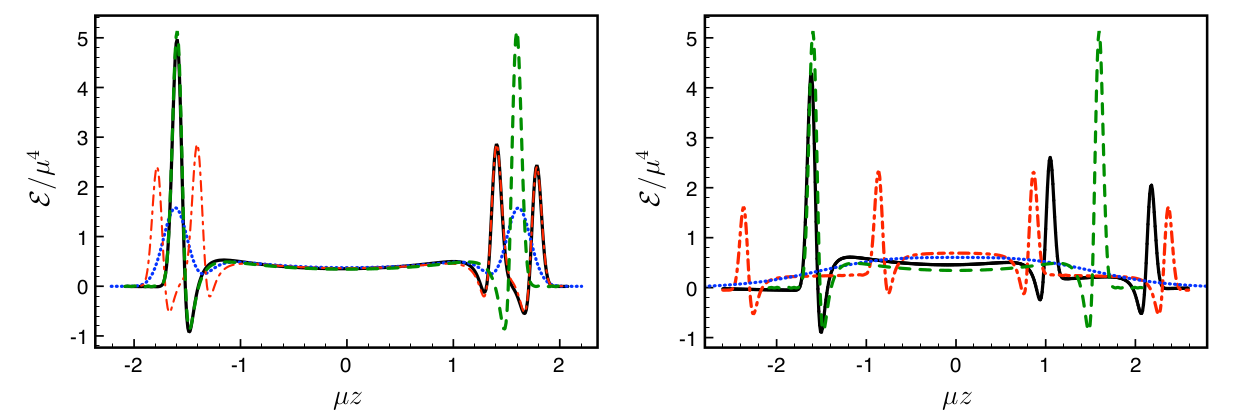}
}
\caption{
Energy profiles after the collision of various superpositions of shock waves. On the left in the coherent regime and on the right in the
incoherent one (figure from \cite{ref18}).}
\label{fig3}
\end{figure}

\begin{figure}[h]
\centerline{
\includegraphics[height=5cm]{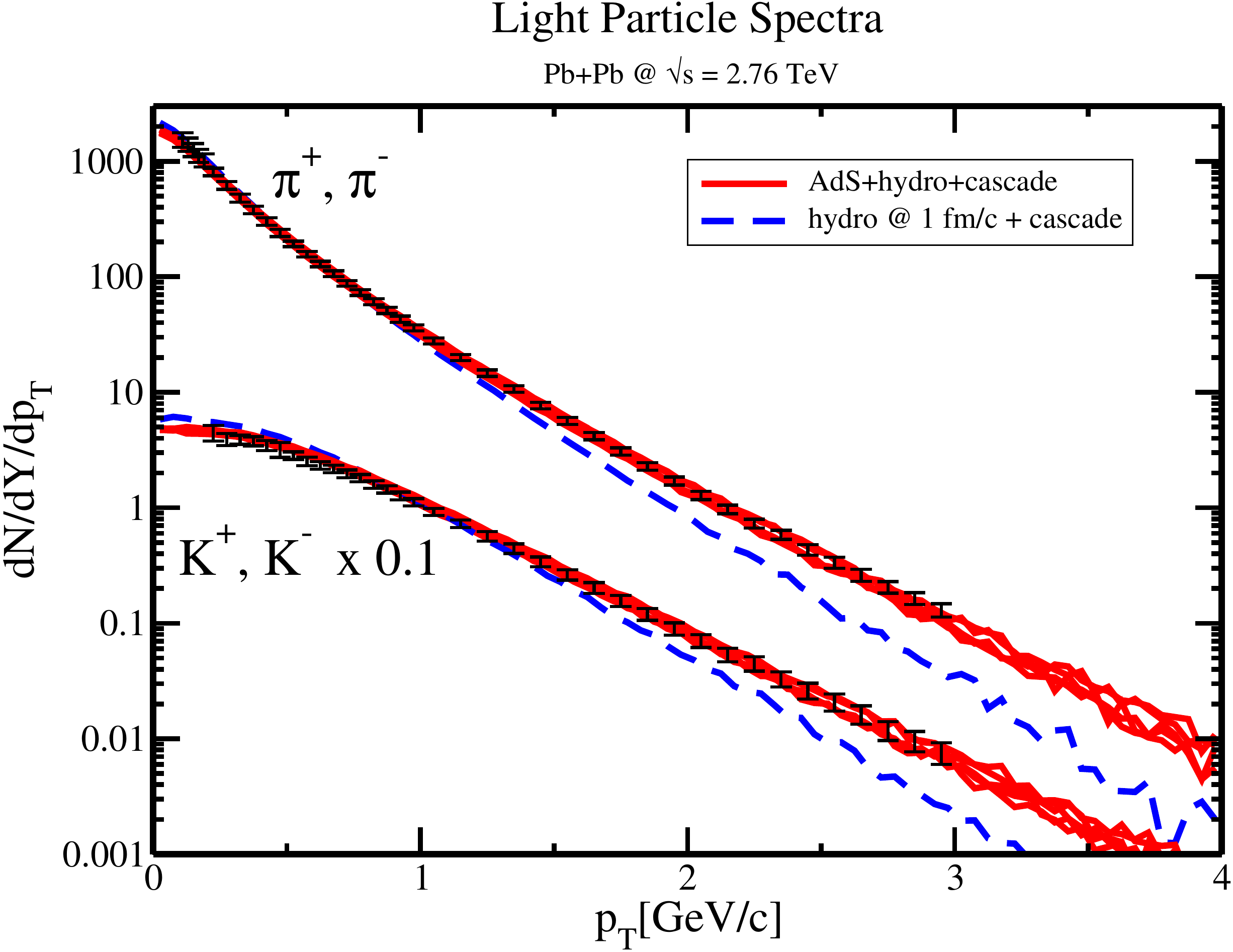}
}
\caption{
A comparison of experimental spectra of light particles produced in central $Pb$-$Pb$ collisions from ALICE with i) the hybrid
AdS/CFT+hydro code and ii) hydro code with conventional initial conditions (figure from \cite{ref20}).}
\label{fig4}
\end{figure}

The chief observation was that the main features of the plasma at midrapidity is insensitive to the structure of the
initial shock waves if the longitudinal separation $L$ along $x^-$ is
\eq
\label{eq12}
L<\f{0.26}{T_{hydr}}
\eqx
as shown in figure 3. On the left hand side we see energy profiles after collisions of three different configurations
of longitudinally separated shock-waves. The separation obeys the above criterion, and the energy profiles in the
central region basically coincide. On the right hand side, the longitudinal separations are larger than in (\ref{eq12}) 
and we see definite differences in the central region.

\subsection{A hybrid study of central nuclear collisions}

The final very interesting work which I would like to mention is a hybrid model of central nuclear collisions
presented in \cite{ref20}. There, nuclei are modelled by a thin shock-wave with realistic radial size. Then, using some
approximations, a numerical relativity simulation of AdS/CFT is used for the pre-equillibrium stage. Then, when
the energy-momentum tensor obtained from the numerical geoemetry becomes well described by hydrodynamics,
the authors extract from it the energy density and flow velocity, and use these quantities as initial conditions for
subsequent hydrodynamic evolution using standard viscous hydrodynamic code for realistic quark-gluon plasma and
kinetic theory for the low-density hadronic stage. The surprising result is that there is an excellent agreement of light
particle spectra with ALICE results (see figure 4).

The key feature provided by the AdS/CFT treatment of the initial nonequilibrium stage is some specific initial
radial flow (‘radial preflow’). The very good agreement with ALICE data suggests that the specific kind of preflow
obtained from the AdS/CFT treatment may be quite realistic. It would be interesting to understand further its main
features.

\section{Conclusions}

The AdS/CFT correspondence provides a very general framework for studying time-dependent dynamical pro-
cesses. What is particularly appealing is that these methods do not presuppose hydrodynamics so are applicable even
to very out-of-equilibrium configurations. Moreover, the treatment of the initial out-of-equilibrium stage and subse-
quent hydrodynamic evolution occurs within the same framework and thus there is no issue with matching together
two different descriptions.

We can get novel qualitative (or even semi-quantitative) insight into the features of the transition to hydrodynamic
behaviour. Indeed, the studies within AdS/CFT, have already uncovered some unexpected generic features.
In particular there is a clear distinction between thermalization and hydrodynamization i.e. a good description
in terms of viscous hydrodynamics. At that transition, the longitudinal and transverse components of the energy-
momentum tensor are still significantly different. Subsequent isotropisation occurs wholly through viscous dissipative
flow. In the boost-invariant setup, there is a simple criterion for the validity of a hydrodynamic description. For all
initial conditions considered, viscous hydrodynamics applies with high accuracy when $\teff \cdot \tau>0.6-0.7$

The AdS/CFT correspondence incorporates in a natural way nonhydrodynamic degrees of freedom in the strongly
coupled plasma. It is also possible to incorporate, in an approximate way, some of these degrees of freedom into a
generalization of a conventional hydrodynamic description.

The AdS/CFT description also exhibits some surprising universal features in shock wave collisions. For thin
shock waves, at the transition to hydrodynamics there is a flat energy profile (in longitudinal coordinate) or equiva-
lently a particular gaussian profile in rapidity. Moreover, this behaviour remains to a large extent independent of the
longitudinal substructure of the colliding shock waves.

Finally, a hybrid study of central heavy-ion collisions used the AdS/CFT analysis as initial conditions for hy-
drodynamics with some specific dynamically generated initial radial preflow. The phenomenological success of this
approach in describing light particle spectra at ALICE is quite surprising and encouraging.

There are still numerous fascinating questions that one could address using the AdS/CFT correspondence. In fact
the possible applications are not restricted just to the physics of thermalization, as was indicated by other AdS/CFT
presentations at this conference \cite{ref21,ref22,ref23,ref24}.

\bigskip

{\bf Acknowledgments.} This work was supported by NCN grant 2012/06/A/ST2/00396.

\end{document}